\documentclass[12pt]{article}
\usepackage{mathrsfs,amsmath,amssymb,amsfonts}

\begin{document}
\title{Non-disturbance criteria of quantum measurements\thanks{This project is supported by Natural Science Foundation of China (10771191 and
10471124) and  Natural Science Foundation of Zhejiang Province of
China (Y6090105).}}
\author{Wu Zhaoqi, Zhang Shifang, Wu Junde\thanks{E-mail: wjd@zju.edu.cn}}
\date{\small Department of Mathematics, Zhejiang University, Hangzhou
310027, P. R. China}
\maketitle
\begin{abstract}
In 2004, Kirkpatrick discussed three ways
(I), (II) and (III) of describing non-disturbance between quantum
measurements $X$ and $Y$, and showed that they are all equivalent to
the compatibility of $X$ and $Y$ if they are both sharp
measurements. In 2005, based on a special sequential product on the
standard effect algebra, Gudder showed that if $X$ and $Y$ are
unsharp measurements, then (I) holds if and only if $X$ and $Y$ are
compatible and $Y$ is sharp measurement; compatibility of $X$ and
$Y$ implies (II), but the converse does not hold, and only (III) is
equivalent to the compatibility of $X$ and $Y$. In 2009, Liu and
Shen and Wu in [J. Phys. A: Math. Theor. {\bf 42}, 185206 (2009), J.
Phys. A: Math. Theor. {\bf 42}, 345203 (2009)] showed that there are
many sequential products on the standard effect algebra. In this
paper, we obtain the same conclusions as Gudder's conclusions for
all these sequential products of the standard effect algebra.
\end{abstract}
~~~~~~~~\small{\textbf{Key Words:} Quantum measurements, non-disturbance,
sequential product}
\section{Introduction}
The fact that quantum measurements can disturb each other is
manifested first by Heisenberg uncertainty principle. This
disturbance is due to the non-commutativity of the position and
momentum operators. In fact, the concepts of non-disturbance,
compatibility, commutativity, coexistence and joint measurability
are closely related to each other and is studied by many authors.\\

In order to state our results, we first need to fix the notations.
Let $H$ be a complex Hilbert space which represents a
quantum-mechanical system $S$. A bounded self-adjoint operator $A$
on $H$ such that $0\leq A\leq I$ is called a {\it quantum effect} on
$H$ (\cite{1,2}). We denote the set of all quantum effects on $H$ by
${\cal E}(H)$ and call it a standard effect algebra, the set of all
orthogonal projection operators on $H$ by $\mathcal {P}(H)$.
Orthogonal projection operators represent sharp yes-no measurements,
while quantum effects represent yes-no measurements that may be
unsharp. Let ${\cal S}(H)$ denote the set of density operators,
i.e., the trace class positive operators on $H$ of unit trace, which
represent the states of quantum system $S$. An {\it operation} is a
positive linear mapping $\Phi: {\cal S}(H)\rightarrow {\cal S}(H)$
such that for each $T\in {\cal S}(H)$, $0\leq tr[\Phi (T)]\leq 1$
([3-5]).\\

Each orthogonal projection operator $P\in {\cal P}(H)$ is associated
with a so-called L\"{u}ders operation $\Phi_L^P: T\rightarrow PTP$,
moreover, when the quantum-mechanical system $S$ is in state $W\in
\mathcal {S}(H)$, the probability that $P$ is observed is given by
$p_{W}(P)=tr(\Phi_L^P(W))=tr(PWP)=tr(PW)$, and the resulting state
after $P$ is observed is $W_{P}=\frac{PWP}{tr(PWP)}$ whenever
$tr(PWP)\neq 0$. If $P, Q$ are two orthogonal projection operators,
then the {\it conditional probability} that $P$ is observed given
that $Q$ has been observed is $p_{W}(P|Q)=\frac{p_W(QPQ)}{p_{W}(Q)}
=\frac{tr(QPQW)}{tr(QWQ)}$ whenever $tr(QWQ)\neq 0.$ These
operations arise in the context of sharp measurements ([4-5]). In
general, each quantum effect $A\in {\cal E}(H)$ gives rise to a
general L\"{u}ders operation $\Phi_L^A: T\rightarrow
A^{\frac{1}{2}}TA^{\frac{1}{2}}$, moreover, when quantum-mechanical
system $S$ is in state $W\in \mathcal {S}(H)$, the probability that
the effect $A$ is observed is given by
$p_{W}(A)=tr(\Phi_L^A(W))=tr(A^{\frac{1}{2}}WA^{\frac{1}{2}})=tr(AW)$,
and the resulting state after $A$ is observed is
$W_{A}=\frac{A^{\frac{1}{2}}WA^{\frac{1}{2}}}{tr(AW)}$ whenever
$tr(AW)\neq 0$. If $A, B$ are two effects, then the conditional
probability that $B$ is observed given that $A$ has been observed is
$p_{W}(B|A)=\frac{p_W(A^{\frac{1}{2}}BA^{\frac{1}{2}})}{p_{W}(A)}
=\frac{tr(A^{\frac{1}{2}}BA^{\frac{1}{2}}W)}{tr(AW)}$ whenever
$tr(AW)\neq 0.$ These operations arise in the context of unsharp
measurements ([4-6]).\\

Let $\Phi_1, \Phi_2$ be two operations. The composition $\Phi_2\cdot
\Phi_1$ is a new operation, called a sequential operation as it is
obtained by performing first $\Phi_1$ and then $\Phi_2$. In general,
$\Phi_2\cdot \Phi_1\neq \Phi_1\cdot \Phi_2$. Note that for any two
quantum effects $A, B\in {\cal E}(H)$ we have $(\Phi_L^B\cdot
\Phi_L^A)=\Phi_L^{A^{\frac{1}{2}}BA^{\frac{1}{2}}}$ $([5,
P_{36-37}])$. This showed that the new quantum effect
$A^{\frac{1}{2}}BA^{\frac{1}{2}}$ yielded by $A$ and $B$ has
important physical meaning, that is,
$A^{\frac{1}{2}}BA^{\frac{1}{2}}$ can be used to describe the effect
by fist measuring $A$ and then measuring $B$. Professor Gudder
called it the sequential product of $A$ and $B$, and denoted it by
$A\circ B$, moreover, $\circ$ has the following properties ([6-8]):\\

(S1). The map $B\rightarrow A\circ B$ is additive for each $A\in
{\cal E}(H)$, that is, if $B+C\leq I$, then $(A\circ B)+(A\circ
C)\leq I$ and $(A\circ B)+(A\circ C)=A\circ (B+C)$.

(S2). $I\circ A=A$ for all $A\in {\cal E}(H)$.

(S3). If $A\circ B=0$, then $A\circ B=B\circ A$.

(S4). If $A\circ B=B\circ A$, then $A\circ (I-B)=(I-B)\circ A$ and
$A\circ (B\circ C)=(A\circ B)\circ C$ for all $C\in {\cal E}(H)$.

(S5). If $C\circ A=A\circ C$, $C\circ B=B\circ C$, then $C\circ
(A\circ B)=(A\circ B)\circ C$ and $C\circ (A+B)=(A+B)\circ C$
whenever $A+B\leq I$.\\

Professor Gudder presented the following open problem in \cite{9}:
Is $A\circ B=A^{\frac{1}{2}}BA^{\frac{1}{2}}$ the only algebraic
operation on ${\cal E}(H)$ which satisfies properties (S1)-(S5)? In
2009, Liu and Wu answered the problem negatively (\cite{10}).\\

We would like to point out that the sequential product $A\circ
B=A^{\frac{1}{2}}BA^{\frac{1}{2}}=A^{\frac{1}{2}}B(A^{\frac{1}{2}})^*$
of $A$ and $B$ can only describe the instantaneous measurement, that
is, the measurement $B$ is completed at once after the measurement
$A$ is performed. In order to describe a more complicated process
where we allow a duration between the measurement $A$ with the
measurement $B$, then we need to replace $A^{\frac{1}{2}}$ with
$f(A)$, $ (A^{\frac{1}{2}})^*$ with $(f(A))^*$, where $f(A)$ is a
function of $A$ which describe the change of $A$ was made by the
duration between $B$ with $A$. Thus, we need to consider the
following general sequential product $f(A)B(f(A))^*$. In order to
guarantee $(f(A))^*=\overline{f}(A)$, we ask $f$ to be a bounded
complex Borel function which is defined on the spectra $sp(A)$ of
$A$.\\

By the above motivation, in \cite{11}, Shen and Wu proved the
following result:\\

\textbf{Theorem 1.1.} Let $H$ be a finite dimensional complex Hilbert
space, $\mathbb{C}$ the set of complex numbers, $\mathbb{R}$ the set
of real numbers, for each $A\in {\cal E}(H)$, $sp(A)$ the spectra of
$A$ and ${\cal B}(sp(A))$ the set of all bounded complex Borel
functions on $sp(A)$. Take $f_A\in {\cal B}(sp(A))$ and $B\in{\cal
E}(H)$, we define

$A\diamond B=f_A(A)B({f_A}(A))^*$.

Then $\diamond$ has properties (S1)-(S5) if and only if the set
$\{f_A\}_{A\in{\cal E}(H)}$ satisfies the following two conditions:

(i) $ \ $  For each $A\in{\cal E}(H)$ and $t\in sp(A)$, $|f_A(t)|=
\sqrt{t}$;

(ii) $ \ $ For any $A,B\in{\cal E}(H)$, if $AB=BA$, then there
exists a complex constant $\xi$ such that $|\xi|=1$ and
$f_A(A)f_B(B)=\xi f_{AB}(AB)$.

Note that for each $A\in{\cal E}(H)$, we can take many  $f_A\in
{\cal B}(sp(A))$ satisfies (i) and (ii), so, Theorem 1.1 showed that
there are many sequential products on ${\cal E}(H)$.

Henceforth, $H$ is always a finite dimensional complex Hilbert space
and the set $\{f_A\}_{A\in{\cal E}(H)}$ satisfies conditions (i) and
(ii).\\

Moreover, in \cite{11}, Shen and Wu still proved that:\\

\textbf{Theorem 1.2.} (1) $ \ $
$f_A(A)\overline{f_A}(A)=\overline{f_A}(A)f_A(A)=A$,
$(f(A))^*=\overline{f}(A)$.

(2) $ \ $ If $0\in sp(A)$, then $f_A(0)=0$.

(3) $ \ $ If $A=\sum\limits^{n}_{k=1}\lambda_{k}E_{k}$, where
$\{E_{k}\}^{n}_{k=1}$ are pairwise orthogonal projections, then
$f_A(A)=\sum\limits^{n}_{k=1}f_A(\lambda_{k})E_{k}$.

(4) $ \ $ For each $E\in {\cal P}(H)$,
$f_E(E)=f_E(0)(I-E)+f_E(1)E=f_E(1)E$.

(5) $ \ $ For any $A,B\in{\cal E}(H)$, $A\diamond B\in{\cal E}(H)$.\\

Take $A\in{\cal E}(H)$ and  $f_A\in \{f_A\}_{A\in{\cal E}(H)}$, we
define $\psi^A: T\rightarrow \overline{f_A}(A)Tf_A(A)$ for $ T\in
{\cal S}(H)$. Now, we can define the probability and conditional
probability which bases on the general sequential product $A\diamond
B=f_A(A)B\overline{f_A}(A)$. For example, when the system is in
state $W\in \mathcal {S}(H)$, the probability that the effect $A\in
{\cal E}(H)$ is observed is given by
$p_{W}(A)=tr(\psi^A(W))=tr(\overline{f_{A}}(A)Wf_{A}(A))$, and the
resulting state after the effect $A$ is observed is
$W_{A}=\frac{\psi^A(W)}{tr(\psi^A(W))}$ whenever $tr(\psi^A(W))\neq
0$, the conditional probability that $B$ is observed given that $A$
has been observed is
$p_{W}(B|A)=p_{W_A}(B)=\frac{tr(f_{A}(A)B\overline{f_{A}}(A)W)}{tr(\psi^A(W))}$
whenever $tr(\psi^A(W))\neq 0.$ Thus, it follows from the definition
and Theorem 1.2 that
\begin{equation}p_{W}(C|A\diamond B)=\frac{p_{(\psi^B\cdot
\psi^A)(W)}(C)}{tr((\psi^B\cdot \psi^A)W)}=
 \frac{tr(C\overline{f_{B}}(B)\overline{f_{A}}(A)Wf_{A}(A)f_{B}(B))}
 {tr(B\overline{f_{A}}(A)Wf_{A}(A))}\end{equation} whenever $tr(B\overline{f_{A}}(A)Wf_{A}(A))\neq 0$.\\

In \cite{12}, Kirkpatrick discussed three ways of describing
non-disturbance between quantum measurements as follows:\\

Let $X$ and $Y$ be two discrete POVMs, i.e., $X=\{A_i\}_{i=1}^m$,
$Y=\{B_j\}_{j=1}^n$, where $A_i, B_j\in {\cal E}(H), i=1, 2, \cdots,
m, j=1, 2, \cdots, n$, and $\sum\limits^{m}_{i=1}A_i=I,
\sum\limits^{n}_{j=1}B_j=I$. Kirkpatrick discussed the following
three ways of describing non-disturbance between quantum
measurements $X$ and $Y$:

(I)  The probability of an established value of $Y$ is
unchanged by the later occurrence of a value of $X$.

(II) The probability of occurrence of a $Y$ value is unchanged
by a preceding execution of $X$.

(III) If $p$ and $q$ are $X$ and $Y$ values, respectively,
then the probability of $p$ followed by $q$ coincides with the
probability of $q$ followed by $p$.\\

Kirkpatrick showed that (I), (II) and (III) are equivalent to the
compatibility of $X$ and $Y$ if they are sharp measurements, i.e.,
when $A_i, B_j\in {\cal P}(H), i=1, 2, \cdots, m, j=1, 2, \cdots,
n$, then (I), (II) and (III) are equivalent to the compatibility of
$X$ and $Y$, that is, $A_iB_j=B_jA_i$ for all $i=1, 2, \cdots, m,
j=1, 2, \ldots, n$.\\

In [13], based on the special sequential product $A\circ B$ of the
standard effect algebra ${\cal E}(H)$, Gudder showed that if $X$ and
$Y$ are unsharp measurements, i.e., when $A_i, B_j\in {\cal E}(H),
i=1, 2, \cdots, m, j=1, 2, \cdots, n$, then (I) holds if and only if
$X$ and $Y$ are compatible and $Y=\{B_j\}_{j=1}^n\subseteq {\cal
P}(H)$; compatibility of $X$ and $Y$ implies (II), but the converse
does not hold, and only (III) is equivalent to the compatibility of
$X$ and $Y$.\\

In this paper, we obtain the same conclusions as Gudder's
conclusions for all the sequential products of the standard effect
algebra ${\cal E}(H)$.

\section{Some Lemmas}
In this section, we present some useful lemmas such that to prove
our main results in section 3.\\

\textbf{Lemma 2.1 (\cite{11}).}  Let $A,B\in{\cal E}(H)$. If $AB=BA$,
then $A\diamond B=B\diamond A=AB$. If $A\diamond B=B\diamond A$ or
$A\diamond B=\overline{f_B}(B)Af_B(B)$, then $AB=BA$.\\

\textbf{Lemma 2.2 ([14 Corollary 4.1.2]).} If $A$ is a normal element
of a $C^{*}$-algebra $\mathscr{U}$, and $A^{k}=0$ for some positive
integer $k$, then $A=0$.\\

\textbf{Lemma 2.3 ([15]).} Let $A\in {\cal B}(H)$ have the following
operator matrix form $$A=\left(
      \begin{array}{cc}
        A_{11} & A_{12} \\
        A_{21} & A_{22} \\
      \end{array}
    \right)
$$
with respect to the space decomposition $H=H_1\oplus H_2$. Then $A\geq 0$ if and only if\\

(1) $A_{ii}\in {\cal B} (H_i)$ and $A_{ii}\geq 0$, $i=1,2$;

(2) $A_{21}=A_{12}^*$;

(3) there exists a linear operator $D$ from $H_2$ into $H_1$ such
that $||D||\leq 1$ and
$A_{12}=A_{11}^{\frac{1}{2}}DA_{22}^{\frac{1}{2}}$.\\

The following lemma is important in establishing the first
non-disturbance criteria.\\

\textbf{Lemma 2.4.} Let $A\in \mathcal {B}(H)$ be a normal operator and
$B\in {\cal E}(H)$. If $AB=BAB$, then $AB=BA$.\\

\textbf{Proof.}\\
\emph{Step 1}. Suppose that $A$ is an invertible
operator. It follows from $AB=BAB$ that $BA^{*}=BA^{*}B$, so
\begin{equation}ABA^{*}=ABA^{*}B.\end{equation}
Since $H$ is a finite dimensional space and $0\leq B \leq I$, by
spectral decomposition, $B$ can be represented as $B=\sum_{i=1}^{n}
\lambda_{i}E_{i},$ where $0\leq \lambda_{i} \leq 1$, $\lambda_{1}
\geq\lambda_{2}\geq \cdots\geq \lambda_{n},$ $\{E_{k}\}^{n}_{k=1}$
is pairwise orthogonal projection operators and $\sum_{k=1}^n=I$.
Thus, we have
$$ABA^{*}E_{i}=\lambda_{i}ABA^{*}E_{i}.$$ That is,
$$(1-\lambda_{i})ABA^{*}E_{i}=0.$$
So, it is easily to obtain that
$$ABA^{*}=ABA^{*}P,$$
where $P$ denotes the orthogonal projection operator corresponding
to the eigenvalue $\lambda_{1}=1$. It follows from $A$ is invertible
that $rank(ABA^{*})=rank(B)$. So from the above equality we can
easily obtain $rank(P)=rank(B)$, which means that $B=P$. By the
condition that $AB=BAB$ and $B=P$ we have
$$AP=PAP.$$
Since $A$ is normal, by functional calculus we can easily get
$A^{*}P=PA^{*}P.$ Thus we obtain $AP=PA$. That is, $AB=BA$.\\

\emph{Step 2.} Suppose that $A\in \mathcal {B}(H)$. Since $dim
H<\infty$, denote $dim H=n$, then $A$ can be represented as
$$A=\left(
       \begin{array}{cc}
        A_1&0\\
        0&A_2\\
       \end{array}
     \right)
$$
with respect to the space decomposition $H=R(A^{n})\oplus N(A^{n})$,
where $A_1$ is an invertible operator and $A_2$ is a nilpotent
operator.\\

By Lemma 2.2, it is easy to see that $A_{2}=0$. And by Lemma 2.3,
$B$ can be represented by
$$B=\left(
       \begin{array}{cc}
        B_{11}&B_{12}\\
        B_{12}^{*}&B_{22}\\
       \end{array}
     \right)
,$$ where $0\leq B_{11}\leq I, 0\leq B_{22}\leq I.$ Then from
$AB=BAB$, we get
$$\left(
       \begin{array}{cc}
        A_1&0\\
        0&0\\
       \end{array}
     \right)\left(
       \begin{array}{cc}
        B_{11}&B_{12}\\
        B_{12}^{*}&B_{22}\\
       \end{array}
     \right)=\left(
       \begin{array}{cc}
        B_{11}&B_{12}\\
        B_{12}^{*}&B_{22}\\
       \end{array}
     \right)\left(
       \begin{array}{cc}
        A_1&0\\
        0&0\\
       \end{array}
     \right)\left(
       \begin{array}{cc}
        B_{11}&B_{12}\\
        B_{12}^{*}&B_{22}\\
       \end{array}
     \right),$$
it implied that
\begin{equation}\left\{\begin{array}{lll}
        A_1B_{11}&=&B_{11}A_1B_{11}\\
        A_1B_{12}&=&B_{11}A_1B_{12}\\
        B_{12}^{*}A_1B_{11}&=&0\\
        B_{12}^{*}A_1B_{11}&=&0
       \end{array}
\right.
\end{equation}

It follows from Step 1 and the first equality that
$A_{1}B_{11}=B_{11}A_{1}$. Then from the third equality we have
$B_{12}^{*}B_{11}A_{1}=0$. Since $A_{1}$ is invertible,
$B_{12}^{*}B_{11}=0$, so $B_{11}B_{12}=0$. Also from the second
equality we have $A_{1}B_{12}=A_{1}B_{11}B_{12}$, thus
$A_{1}B_{12}=0$,  and so $B_{12}=0$. Therefore,
$$B=\left(
       \begin{array}{cc}
        B_{11}&0\\
        0&B_{22}\\
       \end{array}
\right).$$ Hence, we obtain $AB=BA$ in the general case. This
completes the proof.
\section{Non-disturbance criteria}
Our main results are the following:\\

\textbf{Theorem 3.1.} Let $X=\{A_k\}_{k=1}^m$ and $Y=\{B_j\}_{j=1}^n$
be two quantum measurements, where $A_k, B_j\in {\cal E}(H), k=1, 2,
\cdots, m, j=1, 2, \cdots, n$. Then
\begin{equation}p_{W}(B_{j}|B_{j}\diamond A_{k})=1\end{equation} holds for
any $j, k$ and $W\in \mathcal {S}(H)$ if and only if
${A_{k}}{B_{j}}={B_{j}}{A_{k}}$ and $B_{j}\in \mathcal {P}(H)$ for
all $j$.\\

\textbf{Proof.} The sufficiency. By assumption, we have
$\overline{f_{A_{k}}}({A_{k}}){B_{j}}={B_{j}}\overline{f_{A_{k}}}({A_{k}})$,
and $\overline{f_{B_{j}}}(B_{j})=\overline{f_{B_{j}}}(1)B_{j}$
(Theorem 1.2(4)). Thus,
\begin{eqnarray*}p_{W}(B_{j}|B_{j}\diamond A_{k})&=&
\frac{tr(B_{j}\overline{f_{A_k}}(A_{k})\overline{f_{B_j}}(B_j)Wf_{B_{j}}(B_{j})f_{A_{k}}(A_{k}))}
 {tr(A_{k}\overline{f_{B_{j}}}(B_{j})Wf_{B_{j}}(B_{j}))}
\\&=&\frac{tr(\overline{f_{A_{k}}}(A_{k})B_{j}\overline{f_{B_{j}}}(1)B_{j}Wf_{B_j}(B_j)f_{A_k}(A_{k}))}
 {tr(A_{k}\overline{f_{B_{j}}}(B_{j})Wf_{B_{j}}(B_{j}))}
\\&=&\frac{tr(A_{k}\overline{f_{B_{j}}}(1)B_{j}Wf_{B_j}(B_j)}
 {tr(A_{k}\overline{f_{B_{j}}}(1)B_{j}Wf_{B_{j}}(B_{j}))}
\\&=&1.
\end{eqnarray*}
Necessity. Since conditional probability is countably additive in
its first argument, so (4) implies
\begin{equation}p_{W}(B_{i}|B_{j}\diamond A_{k})=0\end{equation}
for $i\neq j$. Thus we have
\begin{equation}\frac{tr(B_{i}\overline{f_{A_{k}}}(A_{k})\overline{f_{B_{j}}}(B_{j})Wf_{B_j}(B_j)f_{A_k}(A_{k}))}
 {tr(A_{k}\overline{f_{B_{j}}}(B_{j})Wf_{B_{j}}(B_{j}))}=0\end{equation}
for all $i\neq j$, whenever
$tr(A_{k}\overline{f_{B_{j}}}(B_{j})Wf_{B_{j}}(B_{j}))\neq 0$. We
can write (6) as
\begin{equation}tr(f_{B_j}(B_j)f_{A_k}(A_{k})B_{i}\overline{f_{A_{k}}}(A_{k})\overline{f_{B_{j}}}(B_{j})W)=0.\end{equation}
Now (7) holds even if
$tr(A_{k}\overline{f_{B_{j}}}(B_{j})Wf_{B_{j}}(B_{j}))= 0$ because
in this case
$$\overline{f_{A_{k}}}(A_{k})\overline{f_{B_{j}}}(B_{j})Wf_{B_j}(B_j)f_{A_k}(A_{k})=0.$$
Since (7) holds for every $W$ we conclude that
\begin{equation}f_{B_j}(B_j)f_{A_k}(A_{k})B_{i}\overline{f_{A_{k}}}(A_{k})\overline{f_{B_{j}}}(B_{j})=0\end{equation}
for all $i\neq j$. We then obtain
$$(B_{i}^{1/2}\overline{f_{A_{k}}}(A_{k})\overline{f_{B_{j}}}(B_{j}))^{*}(B_{i}^{1/2}\overline{f_{A_k}}(A_{k})\overline{f_{B_j}}(B_j))=0$$
for all $i\neq j$. Hence,
$B_{i}^{1/2}\overline{f_{A_{k}}}(A_{k})\overline{f_{B_{j}}}(B_{j})=0$
for all $i\neq j$. So $B_{i}\overline{f_{A_{k}}}(A_{k})B_{j}=0$ for
all $i\neq j$. Summing over $i\neq j$ and using $\sum
\limits_{i}B_{i}=I$, we have
$$0=(I-B_{j})\overline{f_{A_{k}}}(A_{k})B_{j}=\overline{f_{A_{k}}}(A_{k})B_{j}-B_{j}\overline{f_{A_{k}}}(A_{k})B_{j}.$$
Thus
$$\overline{f_{A_{k}}}(A_{k})B_{j}=B_{j}\overline{f_{A_{k}}}(A_{k})B_{j}.$$
Note that $\overline{f_{A_{k}}}(A_{k})$ is a normal operator, by
Lemma 2.4 we obtain that
$$\overline{f_{A_{k}}}(A_{k})B_{j}=B_{j}\overline{f_{A_{k}}}(A_{k}).$$
Taking adjoint, we have
$$B_{j}f_{A_{k}}(A_{k})=f_{A_{k}}(A_{k})B_{j}.$$
Thus, for all $j$ and $k$, we have
$$A_{k}B_{j}=\overline{f_{A_{k}}}(A_{k}){f_{A_{k}}}(A_{k})B_{j}
=\overline{f_{A_{k}}}(A_{k})B_{j}{f_{A_{k}}}(A_{k})
=B_{j}\overline{f_{A_{k}}}(A_{k}){f_{A_{k}}}(A_{k})=B_{j}A_{k}.$$
Now (8) becomes
\begin{equation}A_{k}f_{B_{j}}(B_{j})B_{i}\overline{f_{B_{j}}}(B_{j})=0,\forall
i\neq j.\end{equation} Summing (9) over $k$ gives
$$f_{B_{j}}(B_{j})B_{i}\overline{f_{B_{j}}}(B_{j})=0,\forall
i\neq j.$$ Now summing over $i\neq j$ we have
$$f_{B_{j}}(B_{j})(I-B_{j})\overline{f_{B_{j}}}(B_{j})=0.$$
Hence $B_{j}=f_{B_{j}}(B_{j})B_{j}\overline{f_{B_{j}}}(B_{j})
=f_{B_{j}}(B_{j})\overline{f_{B_{j}}}(B_{j})f_{B_{j}}(B_{j})\overline{f_{B_{j}}}(B_{j})
=B_{j}^{2}.$ That is, $B_{j}\in \mathcal {P}(H)$ for all $j$.\\

\textbf{Theorem 3.2.}  Let $X=\{A_k\}_{k=1}^m$ and $Y=\{B_j\}_{j=1}^n$
be two quantum measurements, where $A_k, B_j\in {\cal E}(H), k=1, 2,
\cdots, m, j=1, 2, \cdots, n$. If $A_{k}B_{j}=B_{j}A_{k}$ for any
$k$ and $j$, then
\begin{equation}p_{W}(B_{j})=\sum \limits_{k} p_{W}(A_{k}\diamond
B_{j})\end{equation} holds for any $j$ and $W\in \mathcal {S}(H)$.\\

\textbf{Proof.}  In terms of traces, (10) becomes
$$tr(\overline{f_{B_{j}}}(B_{j})Wf_{B_{j}}(B_{j}))=\sum \limits_{k}
tr(B_{j}\overline{f_{A_{k}}}(A_{k})Wf_{A_{k}}(A_{k})).$$ That is,
\begin{equation}tr(B_{j}W)=\sum
\limits_{k}
tr(f_{A_{k}}(A_{k})B_{j}\overline{f_{A_{k}}}(A_{k})W).\end{equation}
Since $A_{k}B_{j}=B_{j}A_{k}$, we get
$\overline{f_{A_{k}}}(A_{k})B_{j}=B_{j}\overline{f_{A_{k}}}(A_{k})$.
Then the right side of (11) becomes
\begin{eqnarray*}\sum
 \limits_{k}tr(A_{k}B_{j}W)
&=&tr((\sum \limits_{k}{A_{k}})B_{j}W)
\\&=&tr(B_{j}W).
\end{eqnarray*} So the theorem is proved.\\

Note that the converse of Theorem 3.2 does not hold even for the
special sequential product $A\circ
B=A^{\frac{1}{2}}BA^{\frac{1}{2}}$, so, it does also not hold for
the general sequential product $A\diamond B$.\\

\textbf{Theorem 3.3.} Let $X=\{A_k\}_{k=1}^m$ and $Y=\{B_j\}_{j=1}^n$
be two quantum measurements. Then
\begin{equation}p_{W}(A_{k}\diamond B_{j})=p_{W}(B_{j}\diamond
A_{k})\end{equation} holds for any $k$ and $j$ and $W\in \mathcal
{S}(H)$ if and only if $A_{k}B_{j}=B_{j}A_{k}$ for any $k$ and $j$.\\

\textbf{Proof.}  The sufficiency follows from Lemma 2.1 immediately.
For necessity, in terms of traces, (12) becomes
$$tr(B_{j}\overline{f_{A_{k}}}(A_{k})Wf_{A_{k}}(A_{k}))=
tr(A_{k}\overline{f_{B_{j}}}(B_{j})Wf_{B_{j}}(B_{j})).$$ that is,
\begin{equation}tr(f_{A_{k}}(A_{k})B_{j}\overline{f_{A_{k}}}(A_{k})W)=
tr(f_{B_{j}}(B_{j})A_{k}\overline{f_{B_{j}}}(B_{j})W).\end{equation}
Since (13) holds for all $W$, we have
$$f_{A_{k}}(A_{k})B_{j}\overline{f_{A_{k}}}(A_{k})=
f_{B_{j}}(B_{j})A_{k}\overline{f_{B_{j}}}(B_{j}).$$ That is,
$$A_{k}\diamond B_{j}=B_{j}\diamond A_{k}.$$ 
Then by Lemma 2.1, we obtain
$$A_{k}B_{j}=B_{j}A_{k}.$$ The theorem is proved.\\

\textbf{Remark 1.} For each $E\in {\cal P}(H)$, it follows from Theorem
1.2 that $f_E(E)=f_E(0)(I-E)+f_E(1)E=f_E(1)E$. Using this fact, we
can easily see that if  $P, Q\in {\cal P}(H)$, then
\begin{eqnarray*}P\diamond
Q&=&f_{P}(P)Q\overline{f_{P}}(P)
\\&=&f_{P}(1)PQ\overline f_{P}(1)P
\\&=&PQP=P\circ Q.
\end{eqnarray*}
This showed that if $P$ and $Q$ are two sharp elements, then the
instantaneous measurement and the duration measurement are same.


\begin{thebibliography}{99}

\bibitem{1} Ludwig G. {\it Foundations of Quantum Mechanics (I-II)} (Springer, New York) (1983)

\bibitem{2} Ludwig G. {\it An Axiomatic Basis for Quantum Mechanics(II)} (Springer, New York) (1986)

\bibitem{3} Kraus K. {\it Effects and Operations} (Springer-Verlag, Berlin) (1983)

\bibitem{4} Davies E. B. {\it Quantum Theory of Open Systems} (Academic Press, London) (1976)

\bibitem{5} Busch P., Grabowski M. and Lahti P. J. {\it Operational Quantum Physics} (Springer-Verlag, Beijing World Publishing Corporation) (1999)

\bibitem{6} Gudder S., Nagy G. Sequential quantum measurements. J. Math. Phys. {\bf 42} (2001), 5212-5222.

\bibitem{7} Gudder S., Greechie R. Sequential products on effect algebras. Rep. Math. Phys.  {\bf 49} (2002), 87-111.

\bibitem{8} Gheondea A., Gudder S. Sequential product of quantum effects. Proc. Amer. Math. Soc. {\bf 132} (2004), 503-512.

\bibitem{9} Gudder S. Open problems for sequential effect algebras. Inter. J. Theory. Phys. {\bf 44} (2005), 2219-2230.

\bibitem{10} Liu W. H., Wu J. D. A uniqueness problem of the sequence product on operator effect algebra ${\cal E} (H)$. J. Phys. A: Math. Theor. {\bf 42} (2009), 185206-185215.

\bibitem{11} Shen J., Wu J. D. Sequential product on standard effect algebra ${\cal E} (H)$. J. Phys. A: Math. Theor. {\bf 42} (2009), 345203-345214.

\bibitem{12} Kirkpatrick K. A. Compatibility and probability. arxiv: quant-ph/0403021, 2004.

\bibitem{13} Gudder S. Non-disturbance for fuzzy quantum measurements, Fuzzy Sets and Systems, {\bf 155} (2005), 18-25.

\bibitem{14} Kadison, R., Ringrose, J. Fundamentals of the theory of operator algebras (I, II). American Mathematical Society, New York (1997).

\bibitem{15} Smuljan, J. L. An operator Hellinger integral (Russian). Mat. Sb. (N.S.) {\bf 49} (1959), 381-430.

\end{thebibliography}
\end{document}